\documentstyle[preprint,aps]{revtex}
\tightenlines
\begin{document}

\draft

\title{$\pi$N Elastic Scattering Analyses \\
         and Dispersion Relation Constraints}
\author{Richard A. Arndt and Ron L. Workman}
\address{Department of Physics, Virginia Polytechnic Institute and State
University, Blacksburg, VA 24061}
\author{Igor I. Strakovsky}
\address{Department of Physics, The George Washington University, Washington,
DC 20052}
\author{Marcello M. Pavan}
\address{Massachusetts Institute of Technology, 77 Massachusetts Ave.
Cambridge, MA 02139}

\date{\today}
\maketitle

\begin{abstract}

We present the results of energy-dependent and single-energy
partial-wave analyses of $\pi N$ elastic scattering data with
laboratory kinetic energies below 2.1~GeV.
We have considered the effect of adding
dispersion-relation constraints required for a more reliable
extraction of the $\pi N$ sigma term.  The results of these new
fits are compared with those generated previously, using a reduced
set of constraints, and with a fit which has only utilized
forward constraints
associated with the S-wave scattering lengths.  We compare the
results in terms of their resonance spectra and preferred values
for the $\pi NN$ coupling and $\pi N$ sigma term.

\end{abstract}
\vspace*{0.5in}
\pacs{PACS numbers: 14.20.Gk, 13.30.Eg, 13.75.Gx, 11.80.Et}

\narrowtext
\section{Introduction}
\label{sec:intro}

We have performed partial-wave analyses of pion-nucleon elastic
scattering data up to a laboratory pion kinetic energy of 2.1~GeV.
This work supersedes the analysis~\cite{sm95} (named
SM95) last published by our group. The present analysis (named SP98)
was performed on a larger data base, described in Section~II,
and was constrained by additional fixed-t
dispersion relations beyond those used to generate SM95.
These additional constraints were applied to the $E^{\pm}$ and $A^+$
dispersion relations\cite{Hoehler}, chosen for their connection to
the extraction of the
$\pi N$ sigma term ($\sigma _{\pi N}$).
In Section~III, we will explain how these constraints were added.
For comparison purposes,
a fit to the available data was performed with far fewer dispersion
relation constraints, allowing us to gauge their relative importance.
Results for the baryon spectrum and
associated couplings will be given in Section~IV, along with comparisons
between the solutions SP98 and  SM95.

\section{Database}
\label{sec:dat}

Our previous $\pi N$ scattering analysis\cite{sm95} (SM95)
was based on 10197 $\pi ^{+}p$, 9421 $\pi ^{-}p$, and 1625
charge-exchange data.  Since then, 483 $\pi ^{+}p$, 231
$\pi ^{-}p$, and 48 charge-exchange data
have been added to the database \cite{said}
at energies mainly spanning the $\Delta (1232)$ resonance.

We list these recent (post 1995) additions below.  Some data sets which
we collect are not used in the analyses\cite{flag}, due to
database conflicts, but are retained so that comparisons can be made.  A
complete description of the database and those data not included in our
analyses is available from the authors, and also available through
the interactive program SAID\cite{said}.

Most of the new measurements~\cite{pa97p}-\cite{al95} are from
high-intensity facilities and generally have
small uncertainties.
As mentioned above, a large fraction of the more recent
$\pi ^{\pm}p$ measurements span the $\Delta$
resonance.  These have come mainly from TRIUMF.  From this
source, we have added 106 $\pi ^{+}p$ and 54 $\pi ^{-}p$ differential
cross sections between 140 and 270~MeV\cite{pa97p}.
A further 179 $\pi ^{+}p$ and 51 $\pi ^{-}p$ analyzing power
$A_y$ data between 90 and 270~MeV have been collected using the
new CHAOS facility at TRIUMF\cite{ho97p}.

A few LAMPF, PSI and ITEP
experiments have also been analyzed and added to our database.
The LAMPF additions include 36 high quality polarized charge-exchange
data between 100 and 210 MeV
\cite{ga97p}, 44 $\pi ^{+}p$ and 15 $\pi ^{-}p$ partial total cross
sections between 40 and 280 MeV \cite{kr97p}, and 6 charge-exchange
differential cross sections at 27 MeV\cite{fr97p}.
PSI experiments have provided 5 $\pi ^{-}p$ differential cross sections
at a backward angle between 45 and 70 MeV \cite{jo97} and 11
$A_y$ data at 160 and 240 MeV \cite{ra96}.  After a revised analysis,
the Karlsruhe group, working at PSI, has published a final set of both
$\pi ^{\pm} p$ low-energy differential cross sections \cite{jo95} and
analyzing powers \cite{wi96}.  A final set of polarization parameters
$P$, $R$, and $A$ for $\pi ^{+}p$ at 1300 MeV were contributed by the
ITEP group \cite{al95}.  The distribution of post-1995 data is given
schematically in Fig.~1.

Other experimental efforts will soon provide data in the low to
intermediate energy region.
Precise low-energy measurements of ${\pi}^{-}p$
elastic scattering cross sections\cite{we98}
and $A_y$ data\cite{bi98} have been made at PSI.
$A_y$ data were also
taken between 50 and 100~MeV for both charge channels by CHAOS at
TRIUMF\cite{sm98}.  Further charge-exchange differential cross sections are
expected from a number of labs.  At low energies, the analysis
of data taken at LAMPF between 10 and 40~MeV is close to completion
\cite{is98}.  Above the first resonance region, the Gatchina group
has taken data between 300 and 600 MeV\cite{lo98},
and the Crystal Ball collaboration at BNL has been
collecting data above 500 MeV\cite{br98}.  New
spin-rotation $A$ measurements
are also being carried out at ITEP for ${\pi}^{+}p$ at
1500 MeV\cite{su98}.

\section{Formalism}
\label{sec:form}

There are two main components to the methodology we use in fitting the
$\pi N$ database: the
parameterization scheme and the dispersion-relation constraints.  Our
energy-dependent partial-wave fits are parameterized in terms of a
coupled-channel Chew-Mandelstam K-matrix, as described in Refs.
\cite{sm95,ar85}.  This choice determines the way we
introduce and modify the energy-dependence and account for unitarity
in our fits.  As has been noted in the past, a fit of the K-matrix
elements, expanded in terms of an energy variable, may not result in
a form satisfying all of the requirements imposed by analyticity.
This second part of the analysis is handled iteratively, as has
been described in Ref.\cite{sm95}.  After each iteration, the solution
is checked for compatibility with the imposed set of
dispersion-relation constraints.  A $\chi^2$ penalty due to these
constraints is then added to the data $\chi^2$ in order to force a
best fit to both the data and the constraints.  In our previous
analysis\cite{sm95}, we added constraints from fixed-t dispersion
relations for the $B_{\pm}$ and $C^{\pm}$ invariant amplitudes.  The
$B_{\pm}$ amplitudes are used in the H\"uper plot and are thus tied
to the $\pi NN$ coupling constant.  The $C^{\pm}$ dispersion
relations involve S-wave scattering lengths, and the isospin-odd
combination of these is related to the coupling constant through
the Goldberger-Miyazawa-Oehme sum rule\cite{gmo}.

We have previously made estimates of $\sigma _{\pi N}$
by extrapolating the $A^+$ dispersion relation to the Cheng-Dashen
point\cite{pin10}.
While the $A^+$ dispersion relation was not used as a constraint,
it was found to be reasonably well satisfied by our SM95 solution.
In the present work, we have explicitly constrained this and the
$E^{\pm}$ forward dispersion relations in order to allow a more rigorous
determination of $\sigma_{\pi N}$.

Before giving results, we should recall the importance of the
$\sigma$-term\cite{sainio}, and how it is related to the quantity
we have actually extracted from our fits.
The $\sigma$-term parameterizes the explicit breaking of
chiral symmetry due to the light quark masses,
and may be written as the matrix element
\begin{equation}
  \label{eqn:sigma_def}
 \sigma_{\pi N} = \frac{\hat{m}}{m}
              \frac{\langle m|(\bar{u}u+\bar{d}d-2\bar{s}s)|m\rangle}{1-y}
\end{equation}
where $\hat{m}=(m_u + m_d)/2$ is the average light quark mass, $m$ is
the proton mass, and
\begin{equation}
  \label{eqn:y_{def}}
 y = \frac{2\langle p|\bar{s}s|p\rangle}{\langle p|(\bar{u}u+\bar{d}d)|p\rangle}
\end{equation}
is the strange quark content of the proton.
$\sigma_{\pi N}$ has been calculated theoretically from the SU(3) octet mass
splitting plus meson loop corrections \cite{gl82},
$\sigma_{\pi N}(1-y)$= 35$\pm$5 MeV.

Following the method of Ref.\cite{glls88}, the quantity which we determine
from the data is
\begin{equation}
   \label{bpp_relation}
\Sigma \equiv F^{2}_{\pi}\bar{D}^{+}(\nu=0,t=2m^2_{\pi}) , \\
 \end{equation}
where $F_{\pi}$=92.4 MeV is the pion decay constant, and the $\bar{D}^+$
amplitude is an isoscalar combination of amplitudes, $D = A+\nu B$,
minus the pseudoscalar Born term, with $\nu = (s-u)/4m$.
(The kinematics $(\nu,t )=(0, 2m_{\pi}^2 )$ correspond to the Cheng-Dashen
point.) The difference
$\Sigma - \sigma$ requires theoretical input. Recent estimates\cite{sainio}
suggest a value near 15 MeV.
In using Eq.~(3), we exploit the fact
that $\bar{D}^{+}$ can be  expressed in terms of a power
series\cite{Hoehler},
 \begin{equation}
   \label{eqn:subthresh-expand}
 \bar{D}^{+}(\nu,t) = d^+_{00}+ t\cdot d^+_{01} + \nu^{2}\cdot d^+_{10} + \cdots
 \end{equation}
with coefficients,
$d^{+}_{00}$ and $d^{+}_{01}$,  determined from the
forward dispersion relations for the $C^{+}(t=0)\equiv D^{+}(t=0)$
and $E^{+}\equiv\frac{\partial}{\partial t}(A^{+}+\omega B^{+})|_{t=0}$
amplitudes, respectively. In the expression for $E^+$, $\omega$ denotes
the pion lab energy.
The Sigma term $\Sigma$, which we will quote, is then:
\begin{equation}
   \label{exp_sigma}
   \Sigma = F^2_{\pi}\cdot(d^+_{00}+2m^2_{\pi}\cdot d^+_{01})+\Delta_{D} , \\
\end{equation}
where the ``curvature correction'' term $\Delta_D$
has been estimated\cite{correction} to be $11.9\pm 0.6$ MeV.

The fixed--t $A^{+}$ dispersion relation plays a dual role
in our analysis.  On the
one hand, it provides a constraint at non--zero $t$ which is more
sensitive to the S-waves than the fixed--t $B_{\pm}$ relations used
in our prior analyses\cite{sm95}.  In addition, since
$D\equiv A + \nu B$, the $A^{+}$ subtraction constants
$A^{+}(\nu=0,t)$ are equivalent to $D^{+}(0,t)$.
Thus, linearly extrapolating
these constants (subtracting the pseudovector Born term) to the
Cheng-Dashen point again yields $\Sigma$, providing a
complimentary consistency check to the extraction involving
$d^{+}_{00}$ and $d^{+}_{01}$.

\section{Results and Discussion}
\label{sec:res}

Having fitted the existing database, using various sets of dispersion
relation constraints, we now present and discuss our findings. Most of
the more quantitative results require extrapolations of our amplitudes
into special kinematic regions (for example, in the extraction of
the sigma term and pion-nucleon coupling constant). Further
extrapolations into the complex energy plane are required in order
to determine resonance pole positions and residues.
We can also examine our results from a more qualitative viewpoint,
asking whether the addition of further
constraints has degraded the fit to data.

Fits with different sets of constraints are compared in Table~I.
The SP98 solution has constraints on the $B_{\pm}$ amplitudes (below
800 MeV and for $0\le -t \le 0.3$ (GeV/c)$^2$), the forward $C^{\pm}$
amplitudes (below 600 MeV), the forward $E^{\pm}$ amplitudes (below
600 MeV), and the $A^{+}$ amplitudes (below 900 MeV and for
$0\le -t \le 0.2$ (GeV/c)$^2$).  Our previous solution (SM95) had
constraints on only the $B_{\pm}$ and forward $C^{\pm}$ amplitude,
while the current U372 solution has only forward $C^{\pm}$ constraints.
Clearly, the addition of constraints has not significantly degraded
the fit to data. In all three cases, the overall contribution of
constraint ``data" to the total $\chi ^2$ was far less than the
contribution from cross section and polarization measurements.  (The
$\chi^2$/constraint was generally less than unity.)  The $\pi NN$
coupling constant from these solutions
was also consistent within errors.  For the
solution SP98, the optimal value was 13.72, a result entirely
consistent with that found using SM95 ($g_{\pi NN}^{2}/4\pi$ =
13.75$\pm$0.15\cite{sm95}).
The solution U372 showed more $t$-variation in
extractions via the H\"uper dispersion relation, as should be
expected.  However, the range of values
was similarly consistent with that
found from SM95, which showed negligible $t-$dependence.

Our results for the subthreshold coefficients $d^+_{00}$ and
$d^+_{01}$ are respectively
$-$1.30$m_{\pi}^{-1}$ and 1.27$m_{\pi}^{-1}$.
Defining $\Sigma = \Sigma_d + \Delta_D$,
we find $\Sigma_{d}$=77 MeV.  This result
is consistent with a linear extrapolation of the $A^{+}$ dispersion
relation
subtraction constants, $A^{+}(0,t)$, to the Cheng-Dashen point.
The general agreement of these two determinations
of $\Sigma_{d}$ is significant.  The $A^{+}(0,t)$ subtraction constants
were not fixed {\it a priori}, but determined by insisting that
they should not vary as a function of energy.
In Fig.~2a, we show how well this constraint has been satisfied by SP98.
A set of subtraction constants is plotted for values of $t$ between
0 and $-0.2$ (GeV/c)$^2$. The horizontal lines give an average.
The RMS deviations have been used to estimate the errors plotted in
Fig.~2b. Here a fit of the form,
$A=A_{\rm CD} + A_{\rm lin} (t-2m_{\pi}^2 ) + A_{\rm cusp} B_c(t)$ was used,
where $B_{\rm cusp}$ is a basis function containing a cusp at
$t=4m_{\pi}^2$.  (The terms with coefficients
$A_{\rm lin}$ and $A_{\rm cusp}$ both go to zero (by
construction) at the Cheng-Dashen point.)
The RMS deviations were scaled to give a $\chi^2$/point of unity for the
3 parameter fit. The value of $\Sigma$ thus obtained was 92$\pm$3 MeV.
A linear extrapolation (i.e. $\Sigma_d$)
gave 82 MeV, while a quadratic fit, with the
replacement $B_c(t) \to (t-2m_{\pi}^2)^2$, gave 85 MeV.
This appears to confirm a contribution of 10-15 MeV from the cusp.
The quoted error (3 MeV) is for a fixed value of $g^2/ 4\pi = 13.72$. It
should be noted that, using this method,
$\Sigma = 8.538(A_{\rm CD} - 13.39 g^2/ 4\pi )$ MeV,
and there is a large cancellation between the $A_{\rm CD}$ and $g^2$ terms.
Our value for $\Sigma_d$ is about 50\% larger than the
value\cite{koch_sigma} obtained by the Karlsruhe group.
As a check of our methods, we input the Karlsruhe KA84
phases and reproduced their $\Sigma_{d}$ result exactly.

The partial-wave amplitudes from SM95 and SP98 are displayed in
Fig.~3. Single-energy fits are also plotted here and are compared
to the energy-dependent fit in Table~II. The extracted resonance
parameters, for SP98, are given in Table~III. We find little change
in the I=3/2 partial waves and resonance parameters. The most significant
differences (between SM95 and SP98) are seen in our results for the
low angular momentum I=1/2 states. The largest shifts occur in the
P-waves (P$_{11}$ and P$_{13}$). Of these, the P$_{13}$ resonance pole
has shifted far from its SM95 value,
but is quite weak in our fits. We find a zero
between this pole and the physical energy axis, distorting its effect
and precluding a reliable Breit-Wigner fit.

Two values are given for the $S_{11}$ N(1535) pole position and residue.
The first value was based on our fit to only the elastic $\pi N$ scattering
data bases. The second results when eta-production data are included
in the fit. This fit, and a discussion of its model-dependence, is
given in Ref.\cite{GW}.

\acknowledgments
The authors express their gratitude to B. Bassalleck,  J. T. Brack,
W. J. Briscoe, H. Crannel, J. Comfort, E. Friedman, G. J. Hofman,
C. V. Gaulard, D. Isenhover, M. Janousch, G. Jones,
V. P. Kanavets, A. A. Kulbardis, I. V. Lopatin, M. Mikuz,
W. Plessas, D. Po\v{c}ani\'c, R. A. Ristinen, M. Sadler,
G. R. Smith, V. V. Sumachev, P. Weber, and R. Wieser for providing
experimental data prior to publication or clarification of information
already published.  We also acknowledge useful communications with
G. H\"ohler.
This work was supported in part by the U.~S.~Department of Energy Grants
DE--FG02--97ER41038 and DE--FG02--95ER40901, and a NATO Collaborative
Research Grant 921155U.


\newpage
{\Large\bf Figure captions}\\
\newcounter{fig}
\begin{list}{Figure \arabic{fig}.}
{\usecounter{fig}\setlength{\rightmargin}{\leftmargin}}
\item
{Energy-angle distribution of recent (post-1995) (a) $\pi ^{+}p$, (b)
$\pi ^{-}p$, and (c) charge-exchange (CXS) data.
$\pi ^{+}p$ data are [observable (number of data)]:
d$\sigma$/d$\Omega$~(213),
total elastic cross sections $\sigma ^{t}$~(44),
partial total cross sections (23),
P~(199), R~(2), and A~(2).
$\pi ^{-}p$ data are:
d$\sigma$/d$\Omega$~(159),
total elastic cross sections $\sigma ^{tot}$~(15),
partial total cross sections (6), and
P~(51).
Charge-exchange data are:
d$\sigma$/d$\Omega$~(6),
total cross sections $\sigma ^{tot}$~(6), and
P~(36).  Total cross sections are plotted at zero degrees.}
\item
{(a) Plot of the subtraction constants, from the  $A^+$ dispersion relation,
for $t$-values between 0 and $-0.2$ (GeV/c)$^2$. The
x-marks give values obtained from the solution SP98, while the horizontal
lines give an average for each $t$-value.
(b) Extrapolation of the $A^+(0,t)$ values, from Fig.~2a, to the
Cheng-Dashen point. Errors in the fit are determined from
RMS deviations (see text).
}
\item
{Partial-wave amplitudes (L$_{2I, 2J}$) from 0 to 2.1~GeV.  Solid (dashed)
curves give the real (imaginary) parts of amplitudes corresponding to the
SP98 solution.  The real (imaginary) parts of single-energy solutions are
plotted as filled (open) circles.  The previous SM95 solution \cite{sm95}
is plotted with long dash-dotted (real part) and short dash-dotted
(imaginary part) lines.  The dotted curve gives the value of Im~T - T$^*$~T.
All amplitudes are dimensionless. (a) $S_{11}$, (b) $S_{31}$, (c) $P_{11}$,
(d) $P_{13}$, (e) $P_{31}$, (f) $P_{33}$, (g) $D_{13}$, (h) $D_{15}$,
(i) $D_{33}$, (j) $D_{35}$,  (k) $F_{15}$, (l) $F_{35}$,
(m) $F_{37}$, (n) $G_{17}$, (o) $G_{19}$, (p) $H_{19}$.}
\end{list}

\newpage
\mediumtext
\vfill
\eject
Table~I. Comparison of the solutions SP98, U372, and SM95\cite{sm95}
(see text).  These energy-dependent partial-wave analyses of elastic
$\pi^{\pm} p$ scattering and charge-exchange data had
$N_{prm}$ parameters (I = 1/2 and 3/2) varied in each fit.
\vskip 10pt
\centerline{
\vbox{\offinterlineskip
\hrule
\hrule
\halign{\hfill#\hfill&\qquad\hfill#\hfill&\qquad\hfill#\hfill
&\qquad\hfill#\hfill&\qquad\hfill#\hfill&\qquad\hfill#\hfill\cr
\noalign{\vskip 6pt} %
Solution&T$_{\pi}$~(MeV)&$\chi^2$/$\pi ^{+}p$~data&
$\chi^2$/$\pi ^{-}p$~data&$\chi^2$/CXS~data&$N_{prm}$ \cr
\noalign{\vskip 6pt}
\noalign{\hrule}
\noalign{\vskip 10pt}
SP98 & $0 -2100$ & 22700/10475 & 19515/9531 & 4313/1661 &96/80 \cr
\noalign{\vskip 6pt}
U372 & $0 -2100$ & 22592/10475 & 18151/9531 & 4171/1661 &96/80 \cr
\noalign{\vskip 6pt}
SM95 & $0 -2100$ & 22593/10197 & 18855/9421 & 4442/1625 &94/80 \cr
\noalign{\vskip 6pt}
\noalign{\vskip 10pt}}
\hrule}}
\vfill
\eject
Table~II. Single-energy (binned) fits of combined $\pi ^{\pm}p$
elasic scattering and charge-exchange data, and $\chi^2$ values.
$N_{prm}$ is
the number parameters varied in the single-energy fits, and
$\chi^2_E$ is given by the energy-dependent fit, SP98, over the
same energy interval.
\vskip 10pt
\centerline{
\vbox{\offinterlineskip
\hrule
\hrule
\halign{\hfill#\hfill&\qquad\hfill#\hfill&\qquad\hfill#\hfill
&\qquad\hfill#\hfill&\qquad\hfill#\hfill&\qquad\hfill#\hfill
&\qquad\hfill#\hfill\cr
\noalign{\vskip 6pt} %
T$_{\pi}$~(MeV)&Range~(MeV)&$N_{prm}$&$\chi^2$/data&$\chi^2_E$&&\cr
\noalign{\vskip 6pt}
\noalign{\hrule}
\noalign{\vskip 10pt}
  30 & $  26 -  33 $ &  4 & 203/136 &  248 &&\cr
\noalign{\vskip 6pt}
  47 & $  45 -  49 $ &  4 &  72/81  &  104 &&\cr
\noalign{\vskip 6pt}
  66 & $  61 -  69 $ &  4 & 178/116 &  227 &&\cr
\noalign{\vskip 6pt}
  90 & $  87 -  92 $ &  4 & 126/101 &  141 &&\cr
\noalign{\vskip 6pt}
 112 & $ 107 - 117 $ &  6 &  51/65  &   68 &&\cr
\noalign{\vskip 6pt}
 124 & $ 121 - 126 $ &  6 &  76/60  &   91 &&\cr
\noalign{\vskip 6pt}
 142 & $ 139 - 146 $ &  6 & 173/159 &  201 &&\cr
\noalign{\vskip 6pt}
 170 & $ 165 - 174 $ &  6 & 165/141 &  178 &&\cr
\noalign{\vskip 6pt}
 193 & $ 191 - 194 $ &  6 & 108/107 &  131 &&\cr
\noalign{\vskip 6pt}
 217 & $ 214 - 220 $ &  6 & 116/109 &  137 &&\cr
\noalign{\vskip 6pt}
 238 & $ 235 - 241 $ &  6 & 133/115 &  156 &&\cr
\noalign{\vskip 6pt}
 266 & $ 263 - 271 $ &  6 & 174/123 &  187 &&\cr
\noalign{\vskip 6pt}
 292 & $ 291 - 293 $ &  8 & 141/129 &  197 &&\cr
\noalign{\vskip 6pt}
 309 & $ 306 - 310 $ &  8 & 165/140 &  241 &&\cr
\noalign{\vskip 6pt}
 334 & $ 332 - 335 $ &  8 &  99/59  &  131 &&\cr
\noalign{\vskip 6pt}
 352 & $ 351 - 352 $ &  9 &  81/110 &  123 &&\cr
\noalign{\vskip 6pt}
 389 & $ 387 - 390 $ &  9 &  31/28  &   73 &&\cr
\noalign{\vskip 6pt}
 425 & $ 424 - 425 $ & 10 & 148/139 &  196 &&\cr
\noalign{\vskip 6pt}
 465 & $ 462 - 467 $ & 15 & 351/120 &  431 &&\cr
\noalign{\vskip 6pt}
 500 & $ 499 - 501 $ & 15 & 162/136 &  189 &&\cr
\noalign{\vskip 6pt}
 518 & $ 515 - 520 $ & 17 & 108/79  &  155 &&\cr
\noalign{\vskip 6pt}
 534 & $ 531 - 535 $ & 18 & 133/128 &  179 &&\cr
\noalign{\vskip 6pt}
 560 & $ 557 - 561 $ & 18 & 355/151 &  520 &&\cr
\noalign{\vskip 6pt}
 580 & $ 572 - 590 $ & 18 & 381/286 &  530 &&\cr
\noalign{\vskip 6pt}
 599 & $ 597 - 600 $ & 21 & 258/151 &  458 &&\cr
\noalign{\vskip 6pt}
 625 & $ 622 - 628 $ & 23 & 125/95  &  199 &&\cr
\noalign{\vskip 6pt}
 662 & $ 648 - 675 $ & 23 & 587/352 &  771 &&\cr
\noalign{\vskip 6pt}
 721 & $ 717 - 725 $ & 24 & 209/169 &  286 &&\cr
\noalign{\vskip 10pt}}
\hrule}}
\vfill
\eject
Table~II (continued).
\vskip 10pt
\centerline{
\vbox{\offinterlineskip
\hrule
\hrule
\halign{\hfill#\hfill&\qquad\hfill#\hfill&\qquad\hfill#\hfill
&\qquad\hfill#\hfill&\qquad\hfill#\hfill&\qquad\hfill#\hfill
&\qquad\hfill#\hfill\cr
\noalign{\vskip 6pt} %
T$_{\pi}$~(MeV)&Range~(MeV)&$N_{prm}$&$\chi^2$/$\pi N$~data&$\chi^2_E$&&\cr
\noalign{\vskip 6pt}
\noalign{\hrule}
\noalign{\vskip 10pt}
 745 & $ 743 - 746 $ & 24 & 172/100 &  320 &&\cr
\noalign{\vskip 6pt}
 765 & $ 762 - 767 $ & 25 & 193/169 &  309 &&\cr
\noalign{\vskip 6pt}
 776 & $ 774 - 778 $ & 25 & 226/155 &  294 &&\cr
\noalign{\vskip 6pt}
 795 & $ 793 - 796 $ & 27 & 206/165 &  341 &&\cr
\noalign{\vskip 6pt}
 820 & $ 813 - 827 $ & 29 & 395/304 &  503 &&\cr
\noalign{\vskip 6pt}
 868 & $ 864 - 870 $ & 32 & 283/195 &  418 &&\cr
\noalign{\vskip 6pt}
 888 & $ 886 - 890 $ & 34 & 173/144 &  302 &&\cr
\noalign{\vskip 6pt}
 902 & $ 899 - 905 $ & 35 & 568/416 &  799 &&\cr
\noalign{\vskip 6pt}
 927 & $ 923 - 930 $ & 36 & 232/200 &  354 &&\cr
\noalign{\vskip 6pt}
 962 & $ 953 - 971 $ & 36 & 385/299 &  537 &&\cr
\noalign{\vskip 6pt}
1000 & $ 989 -1015 $ & 38 & 675/423 &  826 &&\cr
\noalign{\vskip 6pt}
1030 & $1022 -1039 $ & 38 & 286/272 &  383 &&\cr
\noalign{\vskip 6pt}
1044 & $1039 -1049 $ & 38 & 365/243 &  486 &&\cr
\noalign{\vskip 6pt}
1076 & $1074 -1078 $ & 43 & 221/218 &  427 &&\cr
\noalign{\vskip 6pt}
1102 & $1099 -1103 $ & 44 & 226/173 &  346 &&\cr
\noalign{\vskip 6pt}
1149 & $1147 -1150 $ & 44 & 327/210 &  450 &&\cr
\noalign{\vskip 6pt}
1178 & $1165 -1192 $ & 44 & 761/394 &  989 &&\cr
\noalign{\vskip 6pt}
1210 & $1203 -1216 $ & 44 & 286/233 &  372 &&\cr
\noalign{\vskip 6pt}
1243 & $1237 -1248 $ & 44 & 455/283 &  650 &&\cr
\noalign{\vskip 6pt}
1321 & $1304 -1337 $ & 44 & 720/401 &  965 &&\cr
\noalign{\vskip 6pt}
1373 & $1371 -1375 $ & 44 & 314/166 &  596 &&\cr
\noalign{\vskip 6pt}
1403 & $1389 -1417 $ & 44 & 549/408 &  822 &&\cr
\noalign{\vskip 6pt}
1458 & $1455 -1460 $ & 46 & 275/258 &  439 &&\cr
\noalign{\vskip 6pt}
1476 & $1466 -1486 $ & 46 & 482/323 &  690 &&\cr
\noalign{\vskip 6pt}
1570 & $1554 -1586 $ & 46 & 839/546 & 1091 &&\cr
\noalign{\vskip 6pt}
1591 & $1575 -1606 $ & 46 & 415/336 &  654 &&\cr
\noalign{\vskip 6pt}
1660 & $1645 -1674 $ & 46 & 552/391 &  809 &&\cr
\noalign{\vskip 6pt}
1720 & $1705 -1734 $ & 46 & 400/279 &  520 &&\cr
\noalign{\vskip 6pt}
1753 & $1739 -1766 $ & 46 & 661/439 &  850 &&\cr
\noalign{\vskip 6pt}
1838 & $1829 -1845 $ & 46 & 456/290 &  746 &&\cr
\noalign{\vskip 10pt}}
\hrule}}
\vfill
\eject
Table~II (continued).
\vskip 10pt
\centerline{
\vbox{\offinterlineskip
\hrule
\hrule
\halign{\hfill#\hfill&\qquad\hfill#\hfill&\qquad\hfill#\hfill
&\qquad\hfill#\hfill&\qquad\hfill#\hfill&\qquad\hfill#\hfill
&\qquad\hfill#\hfill\cr
\noalign{\vskip 6pt} %
T$_{\pi}$~(MeV)&Range~(MeV)&$N_{prm}$&$\chi^2$/$\pi N$~data&$\chi^2_E$&&\cr
\noalign{\vskip 6pt}
\noalign{\hrule}
\noalign{\vskip 10pt}
1875 & $1852 -1897 $ & 46 & 982/682 & 1372 &&\cr
\noalign{\vskip 6pt}
1929 & $1914 -1942 $ & 46 & 852/501 & 1217 &&\cr
\noalign{\vskip 6pt}
1970 & $1962 -1978 $ & 46 & 471/271 &  680 &&\cr
\noalign{\vskip 6pt}
2026 & $2014 -2037 $ & 46 & 398/320 &  695 &&\cr
\noalign{\vskip 10pt}}
\hrule}}
\vfill
\eject
Table~III. Masses ($W_{R}$), half-widths ($\Gamma$/2), and partial
widths for ($\Gamma _{\pi N}$/$\Gamma$) are listed for isospin 1/2
baryon resonances, along with associated pole positions from our
solution SP98~(second sheet poles are denoted by a $\dagger$).
Corresponding residues are given as a modulus and phase (in degrees).
The second set of N(1535) pole parameters, in parenthesis, was found
in Ref.\cite{GW} (see text).
Average values from the Review of Particle Properties\cite{pdg} are
given in square brackets.

\vskip 10pt
\centerline{
\vbox{\offinterlineskip
\hrule
\hrule
\halign{\hfill#\hfill&\qquad\hfill#\hfill&\qquad\hfill#\hfill
&\qquad\hfill#\hfill&\qquad\hfill#\hfill&\qquad\hfill#\hfill
&\qquad\hfill#\hfill\cr
\noalign{\vskip 6pt} %
Resonance &$W_{R}$&$\Gamma$/2&$\Gamma _{\pi N}$/$\Gamma$&Pole &Residue&\cr
\noalign{\vskip 6pt}
(* rating)& (MeV)&   (MeV)  &                 &(MeV)&(MeV, $^{\circ}$)&\cr
\noalign{\vskip 6pt}
\noalign{\hrule}
\noalign{\vskip 10pt}
P$_{11}$(1440)& 1456 &   209    &     0.75           & $1361-i86$ &
(36, -78)&\cr
\noalign{\vskip 6pt}
   & & & & ($1405-i86$)$\dagger$ & (130, -22)$\dagger$&\cr
\noalign{\vskip 6pt}
****          &[1440]&  [175]   &        [0.65]      & & &\cr
\noalign{\vskip 10pt}
D$_{13}$(1520)& 1515 &    46    &         0.65       & $1515-i47$ &
(30, 8) &\cr
\noalign{\vskip 6pt}
****          &[1520]&   [60]   &        [0.55]      & & &\cr
\noalign{\vskip 10pt}
S$_{11}$(1535)& --- &    ---   &         ---       & $1510-i59$ &
(31, -6)&\cr
\noalign{\vskip 6pt}
              &      &          &     &  $(1510-i73)$  &  (40, 7)  &\cr
\noalign{\vskip 6pt}
****          &[1535]&   [75]   &        [0.45]      & & &\cr
\noalign{\vskip 10pt}
S$_{11}$(1650)& 1624 &    49    &         0.78       & $1669-i40$ &
(17, -4)&\cr
\noalign{\vskip 6pt}
****          &[1650]&   [75]   &        [0.72]      & & &\cr
\noalign{\vskip 10pt}
S$_{11}$ &       --- &    ---   &         ---        & $1676-i135$ &
(86, -109)&\cr
\noalign{\vskip 10pt}
D$_{15}$(1675)& 1669 &    75    &         0.40       & $1659-i72$ &
(27, -7) &\cr
\noalign{\vskip 6pt}
****          &[1675]&   [75]   &        [0.45]      & & &\cr
\noalign{\vskip 10pt}
F$_{15}$(1680)& 1679 &    56    &         0.71       & $1674-i59$ &
(40, 6)&\cr
\noalign{\vskip 6pt}
****          &[1680]&   [65]   &        [0.65]      & & &\cr
\noalign{\vskip 10pt}
P$_{11}$(1710)&  --- &   ---    &          ---       & $1697-i135$ &
 (17, 135) &\cr
\noalign{\vskip 6pt}
***          &[1710]&  [50]   &        [0.15]       & & &\cr
\noalign{\vskip 10pt}
P$_{13}$ & --- &   ---         &         ---        & $1515-i129$&
(16, -177)&\cr
\noalign{\vskip 10pt}
F$_{15}$ &  --- &   ---        &           ---       & $1783-i102$ &
(36, -56)&\cr
\noalign{\vskip 10pt}}
\hrule}}
\vfill
\eject
\vfill
\eject
Table~III (continued).
\vskip 10pt
\centerline{
\vbox{\offinterlineskip
\hrule
\hrule
\halign{\hfill#\hfill&\qquad\hfill#\hfill&\qquad\hfill#\hfill
&\qquad\hfill#\hfill&\qquad\hfill#\hfill&\qquad\hfill#\hfill
&\qquad\hfill#\hfill\cr
\noalign{\vskip 6pt} %
Resonance &$W_{R}$&$\Gamma$/2&$\Gamma _{\pi N}$/$\Gamma$&Pole &Residue&\cr
\noalign{\vskip 6pt}
(* rating)& (MeV)&   (MeV)  &                 &(MeV)&(MeV, $^{\circ}$)&\cr
\noalign{\vskip 6pt}
\noalign{\hrule}
\noalign{\vskip 10pt}
G$_{17}$(2190)& 2142 &   235    &         0.25       & $2055-i218$&
(46, -21)&\cr
\noalign{\vskip 6pt}
****          &[2190]&  [225]   &        [0.15]      & & &\cr
\noalign{\vskip 10pt}
H$_{19}$(2220)& 2283 &   201    &         0.26       & $2194-i301$&
(75, -47)&\cr
\noalign{\vskip 6pt}
****          &[2220]&  [200]   &        [0.15]      & & &\cr
\noalign{\vskip 10pt}
G$_{19}$(2250)& 2375 &   404    &         0.12       & --- &
--- &\cr
\noalign{\vskip 6pt}
****          &[2250]&  [200]   &        [0.10]      & & &\cr
\noalign{\vskip 10pt}}
\hrule}}
\vfill
\eject
\vfill
\eject
Table~IV. Parameters for isospin 3/2 baryon resonances.  Notation
as in Table~III.
\vskip 10pt
\centerline{
\vbox{\offinterlineskip
\hrule
\hrule
\halign{\hfill#\hfill&\qquad\hfill#\hfill&\qquad\hfill#\hfill
&\qquad\hfill#\hfill&\qquad\hfill#\hfill&\qquad\hfill#\hfill
&\qquad\hfill#\hfill\cr
\noalign{\vskip 6pt} %
Resonance &$W_{R}$&$\Gamma$/2&$\Gamma _{\pi N}$/$\Gamma$&Pole &Residue&\cr
\noalign{\vskip 6pt}
(* rating)& (MeV)&   (MeV)  &                  &(MeV)&(MeV, $^{\circ}$)&\cr
\noalign{\vskip 6pt}
\noalign{\hrule}
\noalign{\vskip 10pt}
P$_{33}$(1232)& 1234 &    58    &     $\approx$1.0   & $1211-i50$ &
(38, -22)&\cr
\noalign{\vskip 6pt}
****          &[1232]&   [60]   &       [0.994]      & & &\cr
\noalign{\vskip 10pt}
P$_{33}$(1600)&  --- &    ---   &         ---        & $1684-i221$&
(85, 28)&\cr
\noalign{\vskip 6pt}
***           &[1600]&  [175]   &       [0.17]       & & &\cr
\noalign{\vskip 10pt}
S$_{31}$(1620)& 1617 &    56    &        0.29        & $1585-i55$&
(15, -121)&\cr
\noalign{\vskip 6pt}
****          &[1620]&   [75]   &       [0.25]       & & &\cr
\noalign{\vskip 10pt}
D$_{33}$(1700)& 1679 &   142    &        0.16        & $1656-i120$&
(16, -14)&\cr
\noalign{\vskip 6pt}
****          &[1700]&   [150]  &       [0.15]       & & &\cr
\noalign{\vskip 10pt}
F$_{35}$(1905)& 1846 &    141   &        0.12        & $1832-i123$&
(12, -5)&\cr
\noalign{\vskip 6pt}
****          &[1905]&   [175]  &       [0.10]       & & &\cr
\noalign{\vskip 10pt}
P$_{31}$(1910)& 2188  &   489   &        0.27        & $1805-i248$&
(53, -177)&\cr
\noalign{\vskip 6pt}
****          &[1910]&   [125]  &       [0.22]       & & &\cr
\noalign{\vskip 10pt}
D$_{35}$(1930)& 2090 &    294   &        0.11        & $1932-i120$ &
(7, -32)&\cr
\noalign{\vskip 6pt}
***           &[1930]&   [175]  &       [0.15]       & & &\cr
\noalign{\vskip 10pt}
F$_{37}$(1950)& 1923 &    129   &        0.48        & $1881-i118$&
(54, -16)&\cr
\noalign{\vskip 6pt}
****          &[1950]&   [150]  &       [0.38]       & & &\cr
\noalign{\vskip 10pt}}
\hrule}}

\vfill
\eject
\end{document}